\documentclass[global]{svjour}

\usepackage{amsmath,amsfonts,amssymb}
\usepackage{graphicx}

\usepackage[colorlinks,citecolor=blue]{hyperref}
\def\R2Lurl#1#2{\mbox{\href{#1}{\tt #2}}}

\journalname{JCP}

\graphicspath{{.}{figures/}}

\newcommand\eqnref[1]{(\ref{#1})}
\newcommand\figref[1]{Fig.~\ref{#1}}
\newcommand\tableref[1]{Table~\ref{#1}}

\newcommand\sectref[1]{Section~\ref{#1}}

\newcommand{\iu}   {\mathrm{i}}     

\newcommand{\Omegarm}   {\mathrm{\Omega}}

\newcommand{\Psirm}   {\mathrm{\Psi}}
\newcommand{\etal}   {\emph{et al.}~}

\begin{document}
%

\title{Trefftz Difference Schemes on Irregular Stencils}
\author{Igor Tsukerman \email{igor@uakron.edu}}

\institute{Department of Electrical and Computer Engineering, The
University of Akron, OH 44325-3904}

\authorrunning{Tsukerman}

\titlerunning{Flexible Approximation Schemes on Irregular Stencils}
\maketitle

\begin{abstract}
The recently developed Flexible Local Approximation MEthod (FLAME) produces accurate difference schemes
by replacing the usual Taylor expansion with Trefftz  functions -- local solutions of the underlying differential equation.
This paper advances and casts in a general form a significant modification of FLAME proposed recently by Pinheiro \& Webb: a least-squares fit instead of the exact match of the approximate solution at the stencil nodes.
As a consequence of that, FLAME schemes can now be generated on irregular stencils with the number of nodes
substantially greater than the number of approximating functions. The accuracy of the method
is preserved but its robustness is improved. For demonstration, the paper presents a number of numerical examples in 2D and 3D: electrostatic (magnetostatic) particle interactions, scattering
of electromagnetic (acoustic) waves, and wave propagation in a photonic crystal.
The examples explore the role of the grid and stencil size, of the number
of approximating functions, and of the irregularity of the stencils.
\end{abstract}

\begin{keywords}
Flexible local approximation, finite difference schemes, Trefftz functions,
electrostatics, wave propagation, wave scattering, electromagnetic
waves, multiparticle problems, long-range interactions, irregular stencils, meshless methods,
least squares approximation.
\end{keywords}


\section{Introduction}
\label{sec:introduction}
%
Traditional finite difference analysis relies primarily on Taylor expansions.
These are quite general but are accurate only if the underlying solution has the level of smoothness commensurate with the order of the expansion. In particular, the Taylor approximation breaks down
at material interfaces due to jumps in the solution and/or its derivatives.
This leads to the well known ``staircase effect'' in difference schemes
(see a discussion of the algebraic nature of this numerical artifact in
\cite{Tsukerman06,Tsukerman-book07}). Other common situations where the Taylor
series, and hence the corresponding classical schemes, are inaccurate include
boundary layers and sharp peaks or singularities in the vicinity of sources, edges and corners.

The Flexible Local Approximation MEthod (FLAME) \cite{Tsukerman06,Tsukerman-book07}
replaces the Taylor polynomials with much more accurate ``Trefftz'' approximations
by local solutions of the underlying differential equation. As a result,
the approximation accuracy and consequently the consistency error of the scheme
can be improved dramatically.

Previously, FLAME relied on the point-matching of the nodal values of the local approximation.
To fix the key idea, consider a nine-point ($3 \times 3$) scheme for the Laplace equation in 2D.
The solution is approximated locally as a linear combination of eight Trefftz functions,
harmonic polynomials being the most natural choice.\footnote{There are of course
other possibilities: e.g. functions $r^m \exp(\iu m \phi)$ ($m = 0, \pm 1, \pm 2,
\ldots$) in a polar coordinate system $(r, \phi)$ or Green's functions of the form
$|\mathbf{r} - \mathbf{r}_{\mathrm{src}}|^{-1}$, where the source $\mathbf{r}_{\mathrm{src}}$
is located away from the given stencil \cite{Leitao09}, etc.} With nine nodes and only eight free
parameters, the nodal values must be linearly dependent. It is the linear relationship
between them that constitutes the FLAME scheme. Details and a large variety of examples
can be found in \cite{Tsukerman06,Tsukerman-book07,Tsukerman05,Tsukerman-PBG08,Dai-Rubinstein07,Tsukerman-JOPA09};
the applications include electro- and magnetistatics, wave propagation and scattering, the Poisson-Boltzmann equation in macromolecular and colloidal simulation.

One limitation of the point matching procedure is that it links the number of approximating
functions and stencil nodes,
as evident from the above example of eight basis functions for the nine-point stencil.
This connection between functions and nodes has impeded further progress of the method in two different ways.

First, the most natural choices of the stencil and the basis set have not always been feasible.
For instance, consider the standard $3 \times 3$ stencil in 2D problems involving
cylindrical particles. A natural choice of the basis in such cases is cylindrical harmonics \cite{Tsukerman-book07,Tsukerman-PBG08,Tsukerman-JOPA09}
that include the exponential factors $\exp(\iu m \phi)$, $m = 0, \pm 1, \pm 2, \ldots$, where
$\phi$ is the polar angle. Since these harmonics come in pairs $\pm m$ for all $m$ except $m = 0$,
the ``natural'' number of functions in this basis is odd.
At the same time, to obtain a FLAME scheme on
the nine-point stencil, one must choose eight basis functions, which
breaks the natural symmetry: only one index $m$ from the pair $m = \pm 4$ can be taken.
This is more of a nuisance than an actual detriment because all harmonics
up to order $m = 3$ are still included.

A more serious obstacle is that large stencils (in terms of the number of nodes)
require in the established version of FLAME a commensurately large number of
basis functions. Unfortunately, expanded basis sets tend to be poorly conditioned. Yet large stencils are highly desirable in some circumstances, in particular for irregular distributions of nodes where
small stencils can be strongly distorted and unreliable \cite{Dai-adaptive-FLAME08}; including more nodes in the stencil tends to increase robustness.

The disparity between the desirable number of basis functions and stencil nodes became particularly
apparent in three-dimensional electromagnetic vector problems. If the number of basis functions
is commensurate with the number of the nodal degrees of freedom (number of stencil nodes
times three Cartesian components), the basis sets become ill-conditioned.
Pinheiro \& Webb recently overcame this difficulty by using a relatively small
number of approximating functions and applying a least-squares match of the nodal values
\cite{Pinheiro09}. It is this idea that is further advanced in the present paper.

The use of least squares matching allows one to relax or even sever the connection between
the basis functions and the stencil. This facilitates the construction of FLAME on non-canonical irregular stencils.
So far FLAME has mostly been used on regular Cartesian grids  -- not as a requirement but as a practical
matter, to avoid dealing with distorted / skewed stencils. An exception to this practice was
adaptive FLAME in \cite{Dai-adaptive-FLAME08}, where complications due to irregular distributions of nodes in adaptive refinement stencils had to be overcome.

Irregular and adaptive node arrangements are in general highly desirable because the nodes can be concentrated in areas where they are needed the most, in contrast with a regular grid that can only be refined globally. The least squares FLAME described in this paper works on non-canonical stencils more reliably than the previous versions of FLAME.

Irregular stencils are clearly a feature that the new approach, least-squares FLAME, shares with meshless methods (see reviews \cite{Nguyen08,Belytschko94}). Another feature that is shared -- albeit superficially (see below) -- is the least squares matching. I shall, however, refrain from referring to least squares FLAME as a meshless method, for the following reasons.

 The primary and defining feature of FLAME is the accurate Trefftz approximation; ``meshlessness'' is secondary.
 Approaches that have come to be known as ``meshless methods'' are associated with
 quite different approximations. Moving least squares (MLS) techniques,
 the ``reproducing kernel particle method'' (RKPM) \cite{Belytschko94,Krongauz98,Nguyen08},
 and, most recently, maximum-entropy approximations \cite{Sukumar04,Arroyo06}
 are part and parcel of meshless techniques but are quite foreign to FLAME.

Also, the similarity between least squares FLAME and the established meshless methods
does not go very far. MLS, RKPM or maximum-entropy functions in meshless methods are subject
to costly numerical integration and differentiation. In contrast, no integration or differentiation is needed in FLAME at all.

The use of the least squares fit in both MLS and FLAME is also
a superficial similarity. In MLS, the approximation is ``moving,'' in the sense that the
coefficients of the relevant polynomial expansion vary from point to point and are found
via a least squares match at the nodes. The spatial variation of the coefficients
complicates the differentiation and integration that need to be carried out in the numerical procedure.
In FLAME, the approximation is usually non-polynomial
and ``static'': the coefficients are fixed for any given stencil. Further notes on meshless methods can be found in \sectref{sec:Meshless-methods}.

Connections of FLAME with other classes of numerical methods such as variational Trefftz methods,
GFEM, Discontinuous Galerkin,
variational-difference schemes by Moskow \etal \cite{Moskow99}, discontinuous enrichment and others have
previously been discussed in great detail \cite{Tsukerman06}; see in particular Sections 1 \& 2, Fig.~1,
and references in that paper. Here I briefly comment on a few additional contributions, some of them recent.

There is a well established and respectable body of work on special high-order schemes
for various types of problems, in particular the Helmholtz equation \cite{Harari06,Lambe03,Singer06,Baruch09,Nabavi07,Sutmann07}; see Harari's review \cite{Harari06} for further
information and references. There is a commonality of goals but not of the methodology between these techniques and FLAME.

Special difference schemes that can now be viewed as natural particular cases of FLAME have been independently invented by various research groups.
Mei \cite{Mei94} used the fundamental solutions of the Laplace equation to construct
approximate absorbing conditions at the exterior boundary of the computational domain for unbounded problems.
Similar ideas were put forward by Mittra, Boag and Leviatan \cite{Mittra89,Boag94,Boag95}.
Hadley \cite{Hadley02a,Hadley02b} derives difference schemes for the Helmholtz equation
(with applications to electromagnetic waveguide analysis) from the Bessel
function expansion in free space as well as at material boundaries and
corners. Very recently, Chiang and co-authors \cite{Chiang08,Chiang09} developed
high-order schemes for frequency domain analysis of 2-D photonic crystals;
to do so, they carefully match first and higher order derivatives at curved dielectric interfaces.
Also in connection with photonic crystals,
Lu \etal proposed a pseudospectral method of Dirichlet-to-Neumann (DtN) maps \cite{Yuan-Lu06,Hu-Lu08}, where the field is approximated by cylindrical harmonics within the lattice cell and
the Bloch-Floquet conditions are imposed at a set of boundary collocation points.
FLAME, unlike DtN, relies only on \emph{local} analytical approximations over a given stencil,
which tends to be computationally more stable than approximations over the whole lattice cell
\cite{Tsukerman-JOPA09}.

The remainder of the paper is organized as follows.
For completeness, \sectref{sec:Trefftz-FLAME-schemes} summarizes the established FLAME setup.
\sectref{sec:Least-squares-Trefftz-FLAME} presents the new version of FLAME that casts
the idea of Pinheiro \& Webb in a general form. As explained in \sectref{sec:Boundary-conds-FLAME},
boundary conditions in FLAME are easy to impose. \sectref{sec:Consistency-error} proves
that the consistency error in FLAME is commensurate with the approximation accuracy
of the solution by the FLAME bases.

A brief review of meshless methods, in comparison with the least squares FLAME,
is given in \sectref{sec:Meshless-methods}.
A variety of numerical examples in 2D and 3D (\sectref{sec:Num-examples})
include electrostatic (magnetostatic)
particle interactions, scattering of electromagnetic (or acoustic) waves,
and wave propagation in a photonic crystal. The examples also explore the role of the
grid and stencil size, of the number of approximating functions, and of the irregularity of the stencils.
%
\section{Trefftz-FLAME Schemes: A Brief Review}
\label{sec:Trefftz-FLAME-schemes}
%
Trefftz-FLAME is a generalized finite-difference (FD) calculus that
incorporates accurate local approximations of the solution into a
difference scheme. Conceptually, the computational domain
${\Omegarm}$ is covered by a finite number of overlapping subdomains
(``patches'') ${\Omegarm}^{(i)}$, ${\Omegarm} =
{\cup}{\Omegarm}^{(i)}$, $i=1,2,\ldots n$. Each patch contains a
stencil of a global Cartesian grid.

Associated with each patch ${\Omegarm}^{(i)}$ is the local
approximation space
\begin{equation}
    \Psirm^{(i)} ~=~ \mathrm{span} \{\psi_{\alpha}^{(i)}\}, ~~~ \alpha = 1,2,\ldots m
\end{equation}
The number $m$ of approximating functions and the number $M$ of
stencil nodes may in general depend on the patch, but for simplicity
of notation this is not explicitly indicated; i.e. we write $M$, $m$
instead of $M^{(i)}$, $m^{(i)}$. Superscript $(i)$
will occasionally be dropped in other cases as well, when it is clear from
the context that the focus is on a particular patch (stencil) and
there is no possibility of confusion.

The local solution $u_h^{(i)}$ lies
in space $\Psirm^{(i)}$ -- i.e. it is a linear combination of the
local basis functions
\begin{equation}\label{eqn:uh-eq-sum-c-psi}
    u_h^{(i)} ~=~ \sum_{\alpha=1}^m c_{\alpha}^{(i)} \psi_{\alpha}^{(i)}
\end{equation}
In Trefftz-FLAME, the basis functions are chosen to satisfy the
underlying differential equation, along with the interface boundary
conditions.

Clearly, the the nodal values $\underline{u}^{(i)} {\in} \mathbb{R}^M$
of function $u_{h}^{(i)}$ on stencil \#$i$
are linearly related to the coefficient vector
$\underline{c}^{(i)} \equiv \{ c_{\alpha}^{(i)}\} \in \mathbb{R}^m$.
The relevant transformation matrix $N^{(i)}$,
\begin{equation}\label{eqn:u-eq-Nc}
   \underline{u}^{(i)} ~=~ N^{(i)} \underline{c}^{(i)}
\end{equation}
contains the nodal values of the basis functions on the stencil; if
$r_k$ is the position vector of node $k$, then \cite{Tsukerman06,Tsukerman-book07,Tsukerman05}
\begin{equation}\label{eqn:N-matrix}
    N^{(i)} ~~ = ~~ \left(
    \begin{array}{cccc}
      \psi_{1}^{(i)} (r_{1}) & \psi_{2}^{(i)} (r_{1}) & \ldots  & \psi_{m}^{(i)} (r_{1} ) \\
      \psi_{1}^{(i)} (r_{2}) & \psi_{2}^{(i)} (r_{2}) & \ldots  & \psi_{m}^{(i)} (r_{2} ) \\
      \ldots  & \ldots  & \ldots  & \ldots  \\
      \psi_{1}^{(i)} (r_{M}) & \psi_{2}^{(i)} (r_{M}) & \ldots  & \psi_{m}^{(i)} (r_{M} ) \\
    \end{array}
   \right)
\end{equation}
A coefficient vector $\underline{s}^{(i)} \, \in \, R^M$ is
sought to yield the difference scheme
\begin{equation}\label{eqn:scheme-sought}
   \underline{s}^{(i)T} \underline{u}^{(i)} ~=~ 0
\end{equation}
for the nodal values $\underline{u}^{(i)}$ of {\em any} function
$u_{h}^{(i)}$ of form \eqnref{eqn:uh-eq-sum-c-psi}.
Due to \eqnref{eqn:u-eq-Nc} and \eqnref{eqn:scheme-sought},
\begin{equation}\label{eqn:sT-N-c-eq-0}
    \underline{s}^{(i)T} N^{(i)} \underline{c}^{(i)} = 0
\end{equation}
For this to hold for any set of coefficients $\underline{c}^{(i)}$,
one must have
\begin{equation}\label{eqn:s-eq-Null-Nt}
    \underline{s}^{(i)} ~\in~ \mathrm{Null} (N^{(i)T})
\end{equation}
We shall assume that the stencil and the basis set are such that the
null space in the definition of the scheme is of dimension one. If this
happens not to be the case in a particular situation, the stencil and/or
the basis set need to be changed.

For \emph{inhomogeneous} equations (i.e. with a nonzero right hand
side) of the generic form
\begin{equation}
    \mathcal{L} u ~=~ f,
\end{equation}
FLAME schemes are generated by introducing the local splitting of
the solution:
\begin{equation}
    u^{(i)} ~=~ u_0^{(i)} \, + \, u_f^{(i)},
\end{equation}
where $u_0^{(i)}$ is the solution for the homogeneous equation and
$u_f^{(i)}$ is a particular solution of the inhomogeneous equation.
The inhomogeneous FLAME scheme is \cite{Tsukerman06,Tsukerman05}
\begin{equation}\label{eqn:FLAME-scheme-inhomogeneous}
    \underline{s}^{(i)T} \underline{u}^{(i)} ~=~
    \underline{s}^{(i)T} \underline{u}_f^{(i)}
\end{equation}
The coefficients of the scheme and hence the system matrix are the
same for the homogeneous and inhomogeneous problems. The difference
is that, in the presence of sources in the vicinity of a given grid
stencil, the right hand side is formed, as indicated by
\eqnref{eqn:FLAME-scheme-inhomogeneous}, by applying the difference
operator to the nodal values of the particular solution $u_f^{(i)}$.

%
\section{New Schemes: Least Squares Trefftz-FLAME}
\label{sec:Least-squares-Trefftz-FLAME}
%
This section formalizes the modification of FLAME proposed
by Pinheiro \& Webb \cite{Pinheiro09} and casts it in a general form. Specific examples
are presented in the following sections.

One again formally considers a cover ${\cup}{\Omegarm}^{(i)}$, $i=1,2,\ldots n$
of the computational domain by overlapping patches ${\Omegarm}^{(i)}$.
Each patch again contains a set of nodes (stencil). We do not have to assume a regular underlying grid;
the nodes could form fuzzy ``clouds''.

The local approximation spaces are introduced in exactly
the same way as previously; to repeat for easy reference,
\begin{equation}
    \Psirm^{(i)} ~=~ \mathrm{span} \{\psi_{\alpha}^{(i)}\}, ~~~ \alpha = 1,2,\ldots m
\end{equation}
As before, the local solution $u_h^{(i)}$ lies
in $\Psirm^{(i)}$ -- i.e. it is a linear combination of the
local basis functions satisfying the
underlying differential equation, along with the interface boundary
conditions:
\begin{equation}
    u_h^{(i)} ~=~ \sum_{\alpha=1}^m c_{\alpha}^{(i)} \psi_{\alpha}^{(i)}
\end{equation}
At this point, the new approach parts ways with the established version of FLAME
and involves, formally, \emph{two} sets of points. The first one is $n^{(i)}$
stencil nodes $\mathcal{S}^{(i)}$ for which the difference scheme is being generated,
and the other one is a reduced set $\mathcal{S}_1^{(i)} \subset \mathcal{S}^{(i)}$
of $n^{(i)}_1 < n^{(i)}$ nodes at which
least squares matching of the solution is effected, as described below.
In practice, the two sets will typically differ just by
one extra node in $\mathcal{S}^{(i)}$ (see below).


For any given set of nodal values on the reduced set $\mathcal{S}_1^{(i)}$,
the linear relationship \eqnref{eqn:u-eq-Nc} between these values and the expansion coefficients is required to be satisfied in the least squares (l.s.) sense rather than exactly:
\begin{equation}\label{eqn:u-eq-ls-Nc}
   \underline{u}_{S1}^{(i)} ~\overset{l.s.}{=}~ N_1^{(i)} \underline{c}^{(i)}
\end{equation}
Here $N_1^{(i)}$ is the matrix of nodal values of the basis functions on $\mathcal{S}_1^{(i)}$ --
a matrix completely analogous to $N^{(i)}$ of the previous section. The advantage of using the
least squares match is that, as already noted, the connection between the set of basis
functions and the stencil becomes much less rigid than in the established version of FLAME.

More explicitly, the least squares relationship can be written as
\begin{equation}\label{eqn:c-eq-N-plus-u}
   \underline{c}^{(i)} ~=~ N_1^{(i)+} \underline{u}_{S1}^{(i)}
\end{equation}
where $N_1^{(i)+}$ is the generalized inverse (also known as the pseudoinverse or the Moore-Penrose inverse)
\begin{equation}\label{eqn:N-plus-eq-etc}
   N_1^{(i)+} ~=~ \left( N_1^{(i)T} N_1^{(i)} \right)^{-1} N_1^{(i)T}
\end{equation}
assuming that the matrix in the parentheses is nonsingular.
This expression comes directly from the least squares methodology.\footnote{
It is well known that matrix inversion is hardly ever needed in practical computation.
In particular, the pseudoinverse can be computed using the QR method rather than by
explicit matrix inversion.}

The nodal values on the remaining subset of the stencil nodes $S_0^{(i)} = S^{(i)} - S_1^{(i)}$
are required to be \emph{exactly} representable -- as opposed to the least-squares fit --
by a linear combination of the basis functions. Then the Euclidean vector of
these nodal values is
\begin{equation}\label{eqn:u-eq-Nc-eq-N-Kplus-u}
   \underline{u}_{S0}^{(i)} ~=~ N_0^{(i)} \underline{c}^{(i)}
   ~=~ N_0^{(i)} N_1^{(i)+} \underline{u}_{S1}^{(i)}
\end{equation}
where $N_0^{(i)}$, directly analogous to $N_1^{(i)}$, contains the values
of the basis functions on the nodes of $S_0^{(i)}$.
For a homogeneous differential equation (zero right hand side),
we are again looking for a coefficient vector $\underline{s}^{(i)}$
of the difference scheme
\begin{equation}\label{eqn:sT-u-eq-0}
     \underline{s}^{(i)T} \underline{u}^{(i)} ~=~ 0
\end{equation}
or, equivalently due to \eqnref{eqn:u-eq-Nc-eq-N-Kplus-u}
\begin{equation}\label{eqn:sT-NNI-uS1-eq-0}
     \underline{s}^{(i)T}
     \begin{pmatrix}
     N_0^{(i)} N_1^{(i)+} \\
     I_{n1}
     \end{pmatrix}
     \underline{u}_{S1}^{(i)} ~=~ 0
\end{equation}
where $I_{n1}$ is the $n_1 \times n_1$ identity matrix. A natural partitioning
of the difference scheme
$\underline{s}^{(i)T} = (\underline{s}_0^{(i)T}, \, \underline{s}_1^{(i)T})$ is assumed,
with the coefficients $\underline{s}_0^{(i)T}$, $\underline{s}_1^{(i)T}$
corresponding to the nodes in $S_0^{(i)}$ and $S_1^{(i)}$, respectively.

Since the vector of nodal values $\underline{u}_{S1}^{(i)}$
is unknown and in a sense quasi-arbitrary, we require that the above
relationship be satisfied for any such vector. This immediately gives
\begin{equation}\label{eqn:s-eq-Null-N0-N1plus}
    \underline{s}^{(i)} ~\in~ \mathrm{Null}
    \begin{pmatrix}
     N_0^{(i)} N_1^{(i)+} \\
     I_{n1}
     \end{pmatrix}^T
     ~\equiv~
     \mathrm{Null}
    \begin{pmatrix}
     N_1^{(i)+T} N_0^{(i)T},  ~~~  I_{n1}
     \end{pmatrix}
\end{equation}
This is our main expression generalizing the null space condition \eqnref{eqn:s-eq-Null-Nt}
of the established version of Trefftz-FLAME.
We shall always assume that the stencil and the basis are set up in such a way that the
null space in the definition of the scheme is of dimension one. Otherwise
the basis and/or the stencil need to be modified.

It immediately follows from \eqnref{eqn:s-eq-Null-N0-N1plus}
that the scheme can be explicitly written as
\begin{equation}\label{eqn:s-eq-s0-N0N1plus}
    \underline{s}^T ~=~ (s_0^{(i)T}, ~~ -s_0^{(i)T} N_0^{(i)} N_1^{(i)+})
\end{equation}
Typically $S_0^{(i)}$ will contain only one node, in which case, since the null
space is defined up to an arbitrary factor, one can set $s_0^{(i)} = 1$
and obtain
\begin{equation}\label{eqn:s-eq-1-N0N1plus}
    \underline{s}^{(i)T} ~=~ (1, ~~ - N_0^{(i)} N_1^{(i)+})
\end{equation}
This is equivalent to Pinheiro \& Webb's algorithm that they used for electromagnetic
wave scattering in 3D \cite{Pinheiro09}.

The rectangular matrix on the right of \eqnref{eqn:s-eq-Null-N0-N1plus}
has the number of columns equal to the stencil size $n^{(i)}$ and the number of rows
equal to the number of nodes $n_1^{(i)}$ in the substencil $S_1{(i)}$.
Ordinarily, such a matrix will have a one-dimensional nullspace
(leading to a unique FLAME scheme) if the numbers of columns and rows differ by one --
that is, if $n_1^{(i)} = n^{(i)} - 1$ and $S_0{(i)}$ has exactly one node, as stipulated above.

\begin{figure}
\centering
   \includegraphics[width=0.5\linewidth]{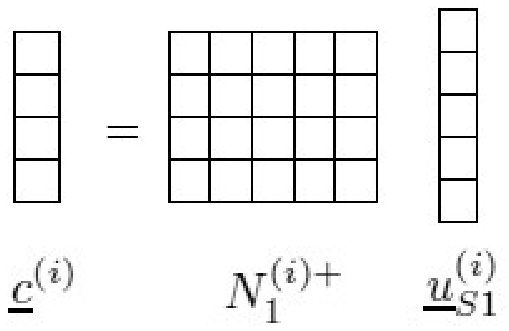}
    \caption{Matrix dimensions in \eqnref{eqn:c-eq-N-plus-u}
    for $n_1^{(i)} = 5$, $m = 4$.}
    \label{fig:c-eq-N1-plus-u-S1}
\end{figure}

\begin{figure}
\centering
   \includegraphics[width=0.9\linewidth]{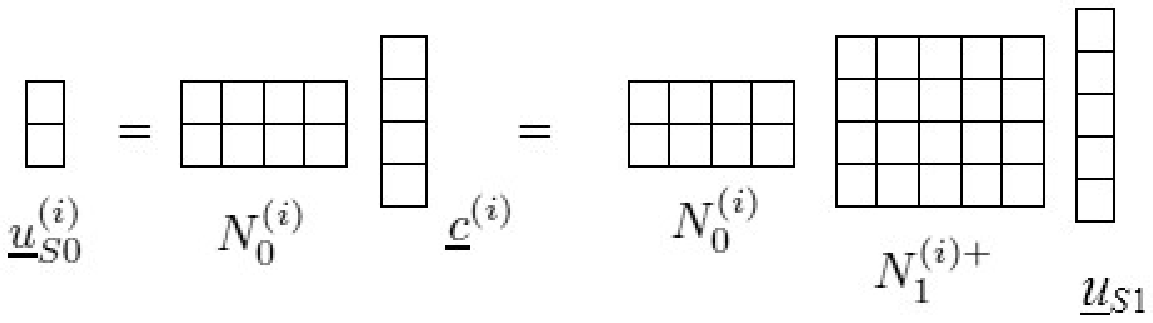}
    \caption{Matrix dimensions in \eqnref{eqn:u-eq-Nc-eq-N-Kplus-u}
    for $n_1^{(i)} = 5$, $n^{(i)} = 7$, $m = 4$.}
    \label{fig:u-S0-eq-N0-c}
\end{figure}

\begin{figure}
\centering
   \includegraphics[width=0.9\linewidth]{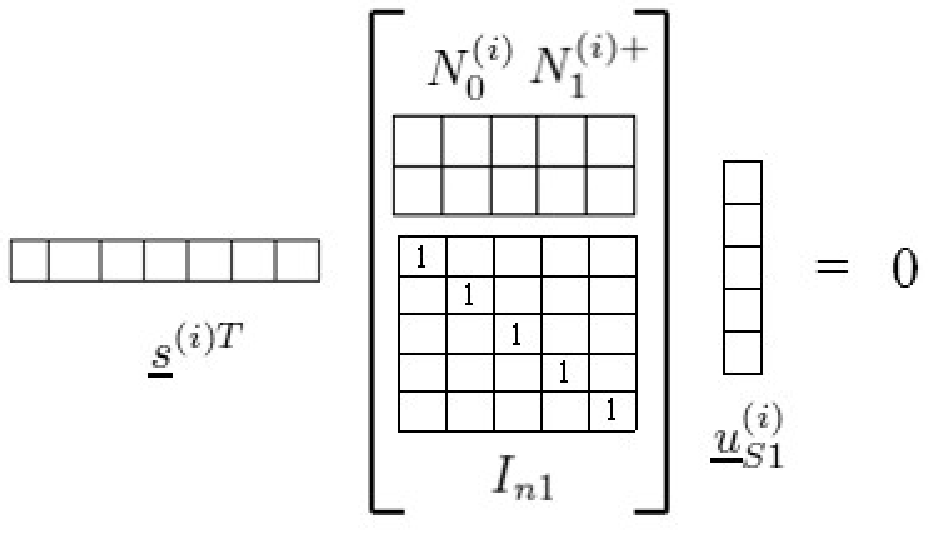}
    \caption{Matrix dimensions and structure in \eqnref{eqn:sT-NNI-uS1-eq-0}
    for $n_1^{(i)} = 5$, $n^{(i)} = 7$.}
    \label{fig:sT-Nu-eq-0}
\end{figure}

\begin{figure}
\centering
   \includegraphics[width=0.65\linewidth]{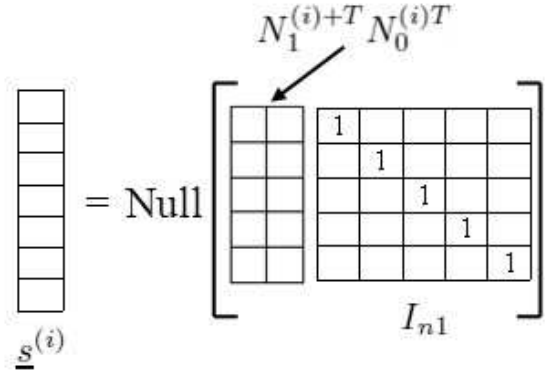}
    \caption{Matrix dimensions and structure in \eqnref{eqn:s-eq-Null-N0-N1plus}
    for $n_1^{(i)} = 5$, $n^{(i)} = 7$.}
    \label{fig:s-eq-Null-NNt}
\end{figure}

For convenience, Figs. \ref{fig:c-eq-N1-plus-u-S1}--\ref{fig:s-eq-Null-NNt} illustrate the dimensions
and composition of the matrices involved in the key equations from \eqnref{eqn:c-eq-N-plus-u} to \eqnref{eqn:s-eq-Null-N0-N1plus}.
In this illustration, there are four approximating functions and seven stencil nodes, five of which
are the ``least squares'' matching nodes of $S_1$.
Although one node rather than two would be typical for $S_0$, two nodes
are assumed, to demonstrate the general principle.

%
\section{Boundary Conditions in FLAME}
\label{sec:Boundary-conds-FLAME}
%
Boundary conditions at the material boundaries
are built into the Trefftz-FLAME bases and therefore need no other
treatment, in sharp contrast with the methods where smooth functions
not obeying the boundary conditions are employed (e.g. MLS or radial basis functions).
Exterior boundary conditions in FLAME do not pose any problem either.
Since all approximations are local, approximations near exterior boundaries
are completely decoupled from approximations elsewhere.  In fact, any
traditional non-FLAME schemes, including perfectly matched layers (PML), can be used at the boundaries.
One could also opt for FLAME schemes at the boundary --
the FLAME PML \cite{Tsukerman-book07,Tsukerman05} being one example.

Natural boundary conditions (Neumann and the like) can either be handled
in the established ways by traditional schemes
or, alternatively, using FLAME with a subset of Trefftz functions satisfying
the natural conditions.

In the customary algorithmic implementation of Dirichlet conditions, one initially disregards the distinction between the nodes at the exterior boundary and the inner nodes. Then, after the system matrix has been
assembled, the usual adjustments are made: terms of the form $a_{ij} u_j$, where $u_j$
is a given Dirichlet value and $a_{ij}$ is a matrix entry, are moved over to the right hand side.
The entry in the right hand side is modified accordingly and the matrix entry $a_{ij}$ is then set to zero.

Even ``fuzzy'' exterior boundaries, where the diffuse set of nodes does not fall exactly
on a predefined surface, could be handled in the same fashion.
There may be some redundancy if the fuzzy boundary layer is slightly ``thicker'' than necessary
in some places. It is trivial but optional to eliminate the redundant nodes (for example, the ones connected only to other Dirichlet nodes) from the system.
%
\section{Consistency Error of FLAME}
\label{sec:Consistency-error}
%
Analysis of the consistency error of least squares FLAME proceeds along the same lines
as in the original version of FLAME \cite{Tsukerman06,Tsukerman-book07}, with
some modifications. Here we shall only consider equations
with the zero right hand side in a given patch (i.e. over a given grid stencil);
the treatment of equations with nonzero r.h.s. is exactly the same as in  the
established versions of FLAME  \cite{Tsukerman06,Tsukerman-book07}.

The consistency error of scheme \eqnref{eqn:sT-u-eq-0}, \eqnref{eqn:s-eq-1-N0N1plus} is,
by definition, obtained by substituting the nodal values of the
exact solution $u^*$ into \eqnref{eqn:sT-u-eq-0}. Let $\epsilon_a
(h)$ be the approximation error of the exact solution $u^{*}$ within a patch
$\Omegarm^{(i)}$:
\begin{equation}\label{eqn:approxmation-error}
    \epsilon_a(h) ~=~ \min_{c^{(i)} \in \mathbb{R}^{m} } \left\| u^{*} -
    \sum\nolimits_{\alpha =1}^{m} c_{\alpha}^{(i)}
    \psi_{\alpha}^{(i)} \right\| _{\infty }
\end{equation}
Equivalently, there exists a coefficient vector $\underline{c}^{(i)}
\, \in \, \mathbb{R}^m$ such that
\begin{equation}\label{eqn:approximation-exact-solution}
    u^{*} ~=~ \sum\nolimits_{\alpha =1}^{m} c_{\alpha}^{(i)}
    \psi _{\alpha }^{(i)} + \eta, ~~~~ \left\| \eta \right\|_{\infty } = \epsilon_a(h)
\end{equation}
within the patch.
For the nodal values, one then has due to \eqnref{eqn:u-eq-Nc-eq-N-Kplus-u}
\begin{equation}\label{eqn:approximation-nodal-vals-exact-sol}
    {\mathcal{N}}^{(i)} u^{*} ~=~ N^{(i)} \underline{c}^{(i)} \,+\, \underline{\eta}
\end{equation}
where ${\mathcal{N}}^{(i)} u^{*}$ and $\underline{\eta}$ = $\mathcal{N}^{(i)}\eta$
are the Euclidean vectors of nodal values of the exact solution and of $\eta$ over stencil $i$,
respectively. $N^{(i)}$ is (as always) the matrix of nodal values of the basis functions. Due to
\eqnref{eqn:approximation-exact-solution},
$$
    \| \underline{\eta} \| _{\infty } ~\leq~ \epsilon_a(h)
$$
In particular, for the nodal values on the ``least squares part'' $S_1^{(i)}$ of the stencil, we have
\begin{equation}\label{eqn:u1-eq-N1c-eta1}
   \underline{u}_1^{*(i)} ~=~ N_1^{(i)} \underline{c}^{(i)} \,+\, \underline{\eta}_1
\end{equation}
Left-multiplying this equation with $N_1^{(i)T}$ and then expressing $\underline{c}^{(i)}$ yields
\begin{equation}\label{eqn:c-eq-N1-plus-h1-minus-eta1}
   \underline{c}^{(i)} ~=~ \left( N_1^{(i)T} N_1^{(i)} \right)^{-1} N_1^{(i)T}
   (\underline{u}_1^{*(i)} \,-\, \underline{\eta}_1)
   ~\equiv~ N_1^{(i)+} (\underline{u}_1^{*(i)} \,-\, \underline{\eta}_1)
\end{equation}
assuming that the matrix in the parentheses is nonsingular.
The $i$th component of the consistency error for scheme \eqnref{eqn:s-eq-1-N0N1plus} is,
by definition,\footnote{Analysis for the slightly more general scheme \eqnref{eqn:s-eq-s0-N0N1plus}
is similar.}
$$
    \left| \epsilon_{ci}(h) \right| ~=~
    \left| \underline{s}^{(i)T} \mathcal{N}^{(i)} u^{*} \right|
    ~=~ \left| u_0^* - N_0^{(i)} N_1^{(i)+} \mathcal{N}_1^{(i)} u_1^* \right|
$$
$$
    =~ \left| (N_0^{(i)} \underline{c}^{(i)} + \eta_0)
    - N_0^{(i)} N_1^{(i)+} (N_1^{(i)} \underline{c}^{(i)} + \underline{\eta}_1) \right|
$$
Now substituting $\underline{c}^{(i)}$ from \eqnref{eqn:c-eq-N1-plus-h1-minus-eta1}, we have
$$
    \left| \epsilon_{ci}(h) \right| ~=~
    \left| \left[ N_0^{(i)} N_1^{(i)+} - N_0^{(i)} N_1^{(i)+} N_1^{(i)} N_1^{(i)+} \right]
    (\underline{u}_1^{*(i)} \,-\, \underline{\eta}_1) \,+\, \eta_0
    \right.
$$
\begin{equation}
    \left. - N_0^{(i)} N_1^{(i)+} \underline{\eta}_1
    \right| ~=~ \left| \eta_0 - N_0^{(i)} N_1^{(i)+}\underline{\eta}_1 \right|
    ~\leq~ | \eta_0 | \,+\, \left\| N_0^{(i)} N_1^{(i)+} \right\| \, \|\underline{\eta}_1 \|
\end{equation}
where the term in the square brackets has in the end vanished because
$A^+ A A^+ = A^+$ for any pseudoinverse matrix $A^+$.
As a result of the generalized inverse falling out of the accuracy estimate,
it can be seen -- somewhat counterintuitively -- that the least squares fit does \emph{not} reduce the accuracy
as compared to the exact nodal match.
The consistency error is still governed primarily by the approximation error.
%
\section{Trefftz-FLAME vs. Meshless Methods}
\label{sec:Meshless-methods}
%
Meshless methods have been developed and studied very extensively
over the last two decades, and excellent contributions and reviews \cite{Belytschko94,Belytschko96,Krongauz98,Liu-Jun-Zhang95,Babuska03},
including a detailed recent one \cite{Nguyen08},
are available. This section is limited to a brief summary of such methods,
with the emphasis on some similarities with FLAME but -- more importantly --
on the differences.

Meshless techniques have two key advantages. One is, by definition, the absence of
complex meshes that may be difficult to generate, especially in 3D, -- although
regular auxiliary grids are still needed to compute the numerical quadratures
in weighted residual and Galerkin methods.

The second advantage is high-order approximation, at least in the regions where the solution
is smooth. This can be accomplished by the Reproducing kernel particle method (RKPM),
Moving least squares (MLS) or similar approximations,
as well as by the recently developed maximum-entropy functions
\cite{Sukumar04,Arroyo06}.

However, meshless methods have problems that are, in a certain sense, an extension
of their advantages. Approximation by smooth functions is accurate in homogeneous
regions but breaks down at interface boundaries where the solution or its derivatives
have jumps. The MLS, RKPM and similar approximating functions and especially their derivatives are too complex to be differentiated or integrated analytically.
Therefore one has to resort to numerical integration and differentiation,
which is in general quite costly. Not only interface boundary conditions
but also the Dirichlet conditions on the exterior boundary of the domain are difficult to impose
numerically because the basis function do not satisfy the Kronecker delta property at the nodes.
The maximum-entropy functions of \cite{Sukumar04,Arroyo06} satisfy a ``weak Kronecker delta''
property due to their rapid decay; however, to compute these functions, one has to
solve a nonlinear optimization problem.

These impediments and ways of getting around them have been
the subject of much research that is still ongoing \cite{Nguyen08}.
The difficulties can be alleviated but not removed. For example,
if collocation instead of weighted residual / Galerkin methods is used,
numerical integration becomes unnecessary but, as an unpleasant trade-off,
higher-order derivatives, commensurate with the order of the differential operator,
of the basis functions have to be calculated. (In the weighted residual methods,
half of the derivatives are usually moved over from the basis functions to
the test functions, thereby reducing the required order of differentiation.
For collocation, this avenue is not available.)

Interface boundary conditions in meshless methods can be treated using Lagrange multipliers
or by introducing additional ``enrichment'' functions. (The latter can also be used
to represent boundary layers, singularities, etc.) This, however, increases the
computational and algorithmic complexity. Lagrange multipliers not only constitute additional unknowns in the
system but may also adversely affect the algebraic properties of the system and may lead to ill-conditioning.
The enrichment functions may have to be mollified by partition of unity factors (to make their support small)
and then are also subject to numerical integration and differentiation.

In contrast, Trefftz-FLAME basis functions satisfy the underlying equation and
the boundary conditions by construction and therefore do not require any additional
constraints to be imposed. In essence, \emph{all} FLAME functions are ``enrichment'' functions
in the sense outlined above; all of them are physical and reflect the local behavior of
the solution. No degrees of freedom are wasted on approximating broader classes of functions
not directly relevant to the problem under consideration.

In short, the difficulties that traditional meshless methods have in dealing with
interface and essential conditions are real, but ways have been developed
to overcome that. FLAME does not have these difficulties to begin with, so there is nothing to overcome.
The price to pay for this advantage is the need to pre-define accurate local analytical
or semi-analytical bases. The complexity of this task depends on the class of problems,
making FLAME quite suitable for some classes and less so for the others.
Examples of problems where the new version of FLAME works well are presented in \sectref{sec:Num-examples};
see also \cite{Tsukerman06,Tsukerman-book07,Tsukerman05,Tsukerman-PBG08} for many other
examples of the established version of FLAME.
%
\section{Computational Examples of Trefftz-FLAME on Irregular Grids}
\label{sec:Num-examples}
%
%

%
\subsection{2D Electromagnetic Wave Scattering}
\label{sec:Num-results-scattering}
%
Let us start by applying least squares FLAME to the classical problem of electromagnetic scattering
from a dielectric cylinder. Non-ideal conductors can be treated in the same manner as dielectrics
if complex permittivity is used. The wave is assumed monochromatic

In the 2D case, where the material parameters and fields are independent of
one coordinate (say, $z$), electromagnetic waves can be decomposed
into two modes. In the $E$-mode (known as the $s$-mode in optics and
frequently, but not always, also referred to as the TM mode) the
electric field has only one component $E = E_z$, whereas the
magnetic field $\mathbf{H}$ has $x$ and $y$ components, with $H_z =
0$. Similarly, for the $H$-mode (also known as the $p$- or TE mode)
$H = H_z$, $E_z = 0$. These modes satisfy the following equations that
are easy to derive from Maxwell's system:
%

\begin{equation}\label{eqn:E-mode}
    \nabla \cdot \mu^{-1} \nabla E ~+~ \omega^2 \varepsilon E \,=\, 0
\end{equation}

\begin{equation}\label{eqn:H-mode}
    \nabla \cdot \epsilon^{-1} \nabla H ~+~ \omega^2 \mu H \,=\, 0
\end{equation}
For numerical testing, let us consider a dielectric cylinder
of radius $r_{\mathrm{cyl}}$ and dielectric permittivity $\epsilon_{\mathrm{cyl}}$.
The host medium is assumed to have the relative dielectric constant of one,
$\epsilon_{\mathrm{out}} = 1$.
The incident field is a plane wave with a wave vector $\mathbf{k}$,
$$
    \mathbf{E}_{\mathrm{inc}} ~=~ \mathbf{E}_0 \exp (\iu \mathbf{k} \cdot \mathbf{r})
$$
under the $\exp(-\iu \omega t)$ complex phasor convention. To complete the formulation,
the standard Sommerfeld radiation conditions can be imposed on the scattered field
$\mathbf{E}_{\mathrm{sc}} = \mathbf{E} - \mathbf{E}_{\mathrm{inc}}$. Numerically,
well established techniques such as PML or absorbing boundary conditions can be
applied on the exterior boundary. Since our focus is on the construction of
accurate difference schemes in the interior, absorbing conditions are a tangential issue,
and in the numerical tests the exact solution was imposed as a Dirichlet condition.
(A textbook solution for a cylindrical scatterer can be found e.g. in \cite{Harrington01}.)
When Dirichlet conditions are imposed, it is tacitly assumed that $\omega$ is not an eigenvalue
of the corresponding operators $\nabla \cdot \mu^{-1} \nabla$ or $\nabla \cdot \epsilon^{-1} \nabla$.

The computational domain is normalized to the unit square $[-0.5, 0.5] \times [-0.5, 0.5]$,
In all numerical experiments, the scatterer is centered at the origin and has the relative permittivity of
$\epsilon_{\mathrm{cyl}} = 9$ and radius $r_{\mathrm{cyl}} = 0.25$.

\begin{figure}
\centering
   \includegraphics[width=0.7\linewidth]{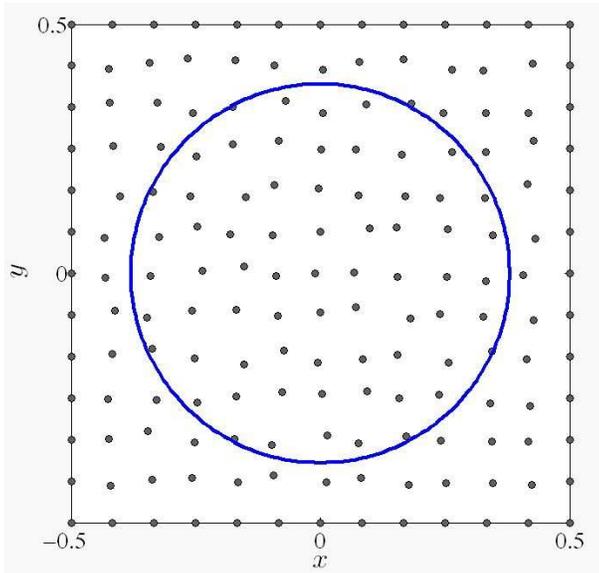}
    \caption{An irregular $12 \times 12$ grid around a cylindrical scatterer.
    ``Fuzziness'' $f = 0.4$.}
    \label{fig:cylinder-fuzzy-grid-12x12}
\end{figure}

Irregular grids are generated by randomly displacing the nodes of a regular Cartesian grid:
\begin{equation}\label{eqn:fuzzy-node-coordinates}
   \hat{x}_i ~=~ x_i \,+\, f \eta_{xi} h_x
\end{equation}
where $x_i$ is the coordinate of the $i$th node of the regular grid,
$h_x$ is the grid size in the $x$-direction,
$\eta_{xi}$ is a random number uniformly distributed within $(-0.5, 0.5)$. Parameter $f$
controls the ``fuzziness'' of the distribution and is chosen between $0$ and $0.5$,
to maintain the majority of the topological connections between each node and its nearest neighbors. Displacements in the $y$ direction are generated in a completely similar way.
\figref{fig:cylinder-fuzzy-grid-12x12} gives an example of a $12 \times 12$ irregular grid.

Construction of the Trefftz bases has been elaborated upon in the previous publications
\cite{Tsukerman06,Tsukerman-book07,Tsukerman05,Tsukerman-PBG08}. Briefly,
the basis functions are chosen as cylindrical harmonics:
$$
    \psi_\alpha^{(i)} ~=~  a_n J_n (k_{\rm cyl} r) \exp(\iu n \phi) ,
    ~~ r \le r_0
$$
$$
    \psi_\alpha^{(i)} ~=~ [b_n J_n(k_{\rm air} r) + H_n^{(1)}
    (k_{\rm air} r)]  \exp(\iu n \phi), ~~ r > r_0
$$
where $J_n$ is the Bessel function, $H_n^{(1)}$ is the Hankel
function of the first kind\footnote{Hankel functions
of the \emph{second} kind should be used if the $\exp(\iu \omega t)$
phasor convention common in electrical engineering is adopted.}, and $a_n$, $b_n$
are coefficients found via the standard conditions on the boundary of the cylinder.
The FLAME basis contains a finite number of harmonics from the monopole
($n$ = 0) up to some highest order $n_{\max}$.

The relative numerical errors for the $E$-mode computed on a $30 \times 30$ FLAME grid, with
a varying number of basis functions, stencil nodes and fuzziness,
are reported in \tableref{table:ls-FLAME-error-scattering-cyl-varying-stencils}.
The stencil of size $m$ for any given node is formally defined as
the set of $m$ nodes closest to that node (including the node itself).
The relative errors were computed as
\begin{equation}\label{eqn:rel-error-scattering-FLAME}
    \delta_\mathrm{rel} ~\equiv~ \frac{ \| \underline{u}_{\mathrm{FLAME}} - \underline{u}_{\mathrm{exact}} \|}
    {\| \underline{u}_{\mathrm{exact}} \| }
\end{equation}
The overall high accuracy of FLAME is evident from the table.
Five to ten correct digits in the value of the field are routinely obtained,
even though the grid is irregular, relatively coarse and \emph{does not represent
the geometry} of the problem in any way. Rather, the geometric and physical information
is built into the FLAME bases.

While the high accuracy was to be expected, there are two surprises in the results
of \tableref{table:ls-FLAME-error-scattering-cyl-varying-stencils}. First,
the fuzziness of the grid does not significantly affect the accuracy --
if anything, the errors on fuzzier grids ($f = 0.2$, $f = 0.4$) tend to be a little lower than on the regular one ($f = 0$).
No ``physical'' reason for that is apparent, especially in light of the fact
that the situation in 3D is different (see the 3D example below).

The second surprise is that, while the accuracy improves as expected
when the number of multipoles $n_{\max}$ is increased from three to four to five,
the relative error goes up from $\sim10^{-11}$ to $\sim10^{-10}$ when
the 6th order multipoles are included. One possible circumstance that might contribute
to this was noted by \v{C}ajko \cite{Cajko08}:
the coefficients of high-order schemes must themselves
be computed with very high precision, to avoid an asymptotic loss of accuracy for fine grids.

\begin{table}
  \centering
  \begin{tabular}{|c|c|c|c|}
    \hline
    $n_{\max}$ (highest harmonic \\in the FLAME basis) & Stencil size & ``Fuzziness factor'' $f$  & Numerical error \\
    \hline
    3 & 9 & 0 & 1.1E-06 \\
    \hline
    4 & 13 & 0 & 2.3E-09 \\
    \hline
    5 & 21 & 0 & 1.29E-11 \\
    \hline
    6 & 21 & 0 & 4.5E-10 \\
    \hline
    \hline
    3 & 9 & 0.2 & 1.1E-06 \\
    \hline
    4 & 13 & 0.2 & 6.2E-11 \\
    \hline
    5 & 21 & 0.2 & 1.5E-11 \\
    \hline
    6 & 21 & 0.2 & 3.6E-10 \\
    \hline
    \hline
    3 & 9 & 0.4 & 8.7E-07 \\
    \hline
    4 & 13 & 0.4 & 7.3E-11 \\
    \hline
    5 & 21 & 0.4 & 1.3E-11 \\
    \hline
    6 & 21 & 0.4 & 4.7E-10 \\
    \hline
  \end{tabular}
  \caption{Relative numerical errors in the nodal values of the electric field
  for least squares FLAME schemes with varying bases and stencils
  on a $30 \times 30$ grid. Scattering from a dielectric cylinder with $\epsilon_{\mathrm{cyl}} = 9$,
  $r_{\mathrm{cyl}} = 0.25$.}
  \label{table:ls-FLAME-error-scattering-cyl-varying-stencils}
\end{table}

%
\subsection{Wave Propagation in Photonic Crystals}
\label{sec:Photonic-crystals}
%
Photonic crystals (PhC) are artificial periodic structures that exhibit very
peculiar characteristics of wave propagation (e.g. the photonic bandgap
\cite{Bykov72,Bykov75,Yablonovitch87,John87}) and are finding
various applications in lightwave technology \cite{Johnson01,Sakoda05,Tsukerman-book07}.
Our goal here is limited to using PhC as a demonstration example for the new version
of FLAME.

\begin{figure}
\begin{center}
   \includegraphics[width=0.75\linewidth]{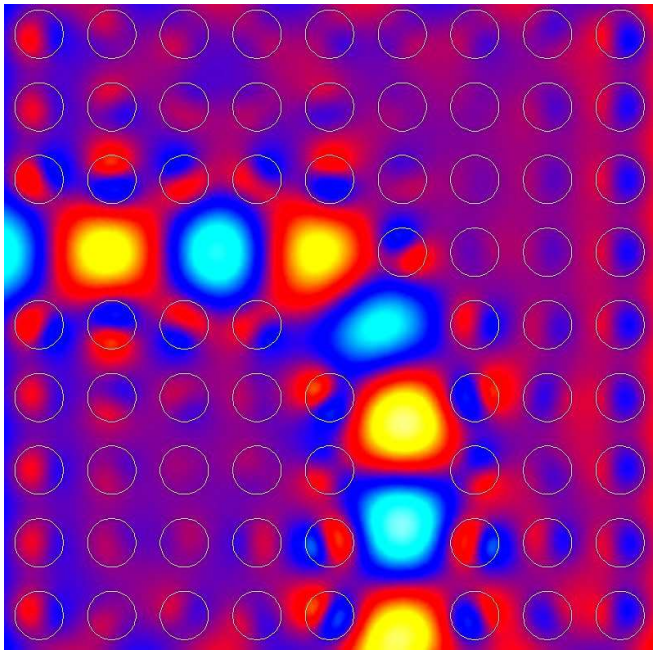}
    \caption{The imaginary part of the electric field in the photonic
    crystal waveguide bend. The real part looks qualitatively similar.
    (Reprinted by permission from \cite{Tsukerman05} \copyright 2005 IEEE.)}
    \label{fig:wave-in-photonic-crystal}
\end{center}
\end{figure}

For convenience of comparison, consider the same numerical example as in \cite{Tsukerman-book07,Tsukerman05}:
a photonic crystal due to Fujisawa \& Koshiba \cite{Fujisawa04},
with a square lattice of cylindrical coaxial dielectric rods and a bended waveguide
obtained by eliminating a few of these rods (\figref{fig:wave-in-photonic-crystal}).
The $E$- and $H$-modes of the wave are governed by equations \eqnref{eqn:E-mode} and
\eqnref{eqn:H-mode}, respectively.

As in the previous tests \cite{Tsukerman-book07,Tsukerman05},
the dielectric constant is set to $\epsilon_\mathrm{rod} = 9$ for the rods
(index of refraction $n = 3$) and to unity for the outside medium. The radius of the cylinders and the
wavenumber are normalized to unity; the air gap between the
neighboring rods is equal to their radius. The field distribution is
shown in \figref{fig:wave-in-photonic-crystal} for illustration. Simplified
boundary conditions are set equal to an externally applied plane wave.
This is adequate for the demonstration example; conditions at the ports of the guide
are a separate issue, and a way to handle them in FLAME was developed by Pinheiro \& Webb
\cite{Pinheiro07}.

An excellent agreement between the established version of Trefftz-FLAME
and independent FEM simulations has been previously demonstrated
\cite{Tsukerman-book07,Tsukerman05}; in fact, numerical evidence
was convincing that the numerical error was in this case
orders of magnitude lower in FLAME than in FEM with a comparable number
of unknowns.

\begin{figure}
  \begin{center}
  \includegraphics[width=\linewidth]{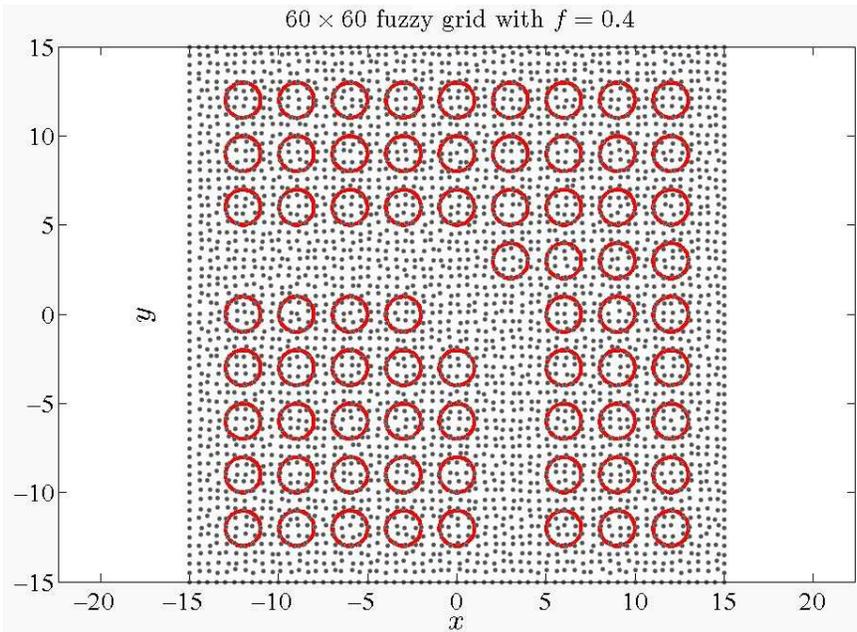}\\
  \caption{A $60 \times 60$ irregular FLAME grid in the photonic crystal example. ``Fuzziness factor'' $f = 0.4$.}\label{fig:PhC-fuzzy-grid-60x60}
  \end{center}
\end{figure}

Our goal here is to compare the least squares FLAME on fuzzy
node clouds with the previous Trefftz-FLAME on regular Cartesian grids,
thereby exploring the influence of the ``fuzziness'' on the numerical accuracy.
%
%
The ``fuzzy'' node coordinates $\hat{x}_i$, $\hat{y}_i$ are generated as
in \eqnref{eqn:fuzzy-node-coordinates}. \figref{fig:PhC-fuzzy-grid-60x60} shows an example of an irregular $60 \times 60$ FLAME grid
with the ``fuzziness factor'' $f = 0.4$.

\begin{figure}
  \begin{center}
  \includegraphics[width=\linewidth]{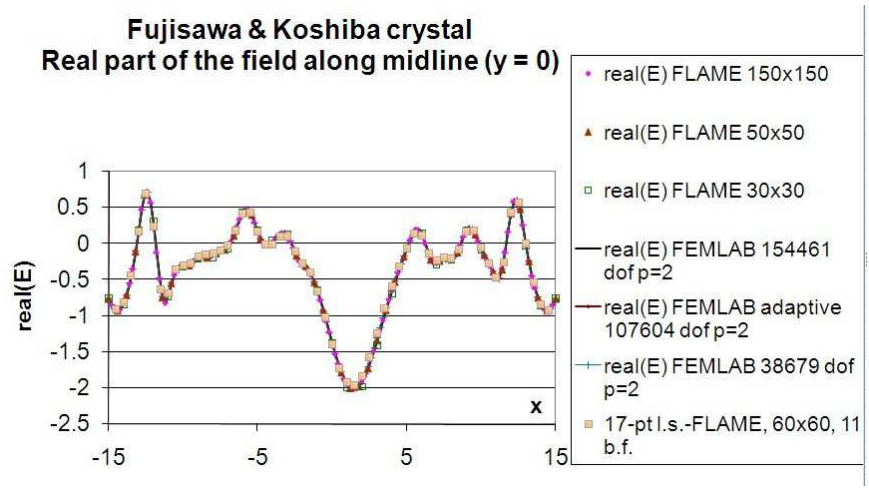}\\
  \caption{Field distribution in the Fujisawa--Koshiba photonic
  crystal along the central line $y$ = 0. Solution by the new version of FLAME
  coincides with the previous FEM and FLAME results, despite the quite irregular
  distribution of nodes in the new test. For testing, least-squares FLAME
  was applied on 17-point stencils with 11 basis functions.}\label{fig:PhC-E-vs-x-midline}
  \end{center}
\end{figure}

This grid was used to produce the least squares FLAME results in
\figref{fig:PhC-E-vs-x-midline}, where the distribution of the electric field along
the midline of the crystal ($y = 0$) is shown for the $E$-mode. The result obtained with the new version of FLAME  coincides with the various results obtained previously with FEM and FLAME, despite the quite irregular
distribution of nodes in the new test (see \figref{fig:PhC-fuzzy-grid-60x60}).

Note that for the $60 \times 60$ grid there are about 12.5 points
per wavelength (ppw) in the air but only 4.2 ppw in the rods, and
yet the FLAME results are very accurate even on an irregular grid, due to the Trefftz
approximation. Alternative methods that
rely on generic polynomial approximations would require
a substantially higher number of ppw to achieve the same accuracy.

The FLAME field values plotted in \figref{fig:PhC-E-vs-x-midline} were produced
by interpolation, similar to the way described in \cite{Tsukerman-book07}, pp.~290--291,
for the previous version of FLAME. Namely, for a given set of nodal values on a stencil,
one finds the Trefftz expansion coefficients $c^{(i)}$  by solving the least squares problem
either on the substencil $S_1$ \eqnref{eqn:u-eq-ls-Nc} or, alternatively, on the whole stencil.
The interpolation is then effected by the Trefftz expansion \eqnref{eqn:uh-eq-sum-c-psi}.
%
\subsection{Electrostatic Particle Interactions in 3D}\label{sec:Num-results-3D-electrostatics}
%
\subsubsection{Formulation and setup.}
The 3D numerical example, chosen to be simple enough for illustration
but extendable to more complex cases, involves two spherical dielectric particles in a
uniform external electrostatic field.

A uniform field induces only
the dipole mode in a \emph{single} particle \cite{Harrington01}, but for two or more
particles higher-order multipoles do arise. A semi-analytical solution
can be obtained by the well known multipole-multicenter method (mmc)
\cite{Cheng99,Mishchenko02,Tsukerman-book07}, where the potential is sought as a superposition of multipole
expansions centered on each of the particles. To impose the boundary
conditions on any given particle, one needs to translate the multipole
expansions from all other particles to that particle; this is accomplished
by the standard multipole translation formulas (e.g. \cite{Cheng99}).
Since the details of this procedure are easily found in the literature and
are tangential to the subject matter of this paper, they are omitted here.

The electrostatic equation in terms of the total electric potential $u$
is\footnote{$u$ is used instead of the somewhat more conventional $\phi$ to
avoid confusion with the polar coordinate.}
\begin{equation}\label{eqn:div-eps-grad-phi-eq-0}
    \nabla \cdot \epsilon \nabla u ~=~ 0;
    ~~~ u(\mathbf{r}) - u_\mathrm{ext}(\mathbf{r}) \rightarrow 0
    ~~\mathrm{as} ~~ r \rightarrow \infty
\end{equation}
where $u_\mathrm{ext}(\mathbf{r})$ is the applied (external) potential,
which for a uniform field is a linear function of coordinates. If understood
in the sense of distributions, \eqnref{eqn:div-eps-grad-phi-eq-0} implicitly
includes the standard conditions on interface boundaries: the continuity
of the potential and the normal component of the electric flux density
$\mathbf{D} = \epsilon \mathbf{E}$. The div-grad equation \eqnref{eqn:div-eps-grad-phi-eq-0}
can also be rewritten in terms of the ``scattered'' potential
$u_s \equiv u - u_\mathrm{ext}$, but there is no particular need to do so here.

The construction of FLAME bases is analogous to the cylindrical case
and involves spherical harmonics defined inside and outside a given particle
and matched at its interface. Details can be found in \cite{Tsukerman05,Tsukerman-book07}.
It is possible to increase the accuracy of FLAME by using more complex
mmc bases over several neighboring particles as done in \cite{Dai-adaptive-FLAME08};
however, in this paper single-center basis functions are used for the sake of practical simplicity.

The mmc expansion with a large number of harmonics
(64 terms per particle in the expansion) is used
to compute a quasi-analytical solution for reference and error analysis.

The two spherical particles in our 3D example have radii $r_1 = 1$ and $r_2 = 1.25$
and are centered on the $z$ axis at points $z_1 = -2.5$, $z_2 = 2$, respectively.
The lack of geometric symmetry is deliberate, to avoid any computationally favorable
artifacts due to symmetry. The dielectric constants inside and outside the particles
in all numerical experiments were chosen as $\epsilon_{\mathrm{in}} = 5$, $\epsilon_{\mathrm{out}} = 1$,
respectively. A cubic computational domain of size $L_{\mathrm{dom}} = 10$ centered at the origin
was used.
%
\subsubsection{Regular Cartesian grids.}
%
The first set of tests is performed on regular Cartesian grids. The relative simplicity
of this example notwithstanding, note that the regular grids do \emph{not} carry
any accurate geometric information about the shapes of the particles; that information
is implicitly imbedded in the FLAME approximating functions. In contrast with the
established version of FLAME, now the stencil size is to a large degree independent
of the number of basis functions.

\begin{figure}
  \centering
  \includegraphics[width=0.9\linewidth]{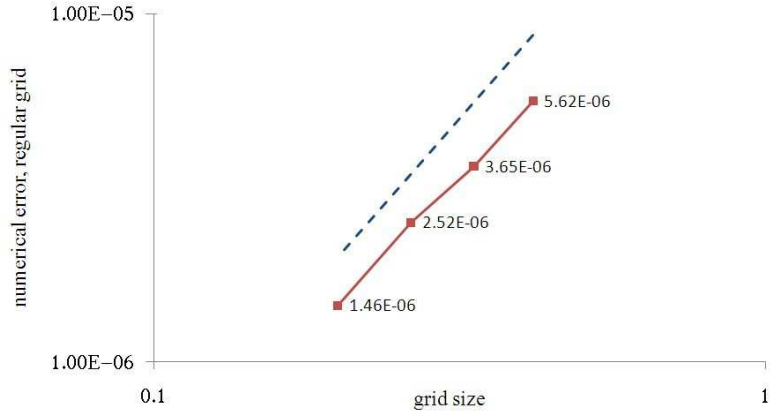}
  \caption{Relative error in potential, regular grids.
  19-point least-squares FLAME with 9 basis functions.
  Test for two spherical particles. To evaluate the numerical error,
  the mmc solution with 64 terms per particle was taken as quasi-exact.
  The dashed line is a visual aid indicating the $\mathcal{O}(h^2)$ convergence rate.}
  \label{fig:error-vs-h-reg-grids-two-spheres}
\end{figure}

\figref{fig:error-vs-h-reg-grids-two-spheres}
illustrates convergence of 19-point least squares FLAME with nine basis functions.
The relative error plotted in the figure was calculated as
\begin{equation}\label{eqn:relative-error_FLAME}
    \delta_\mathrm{rel} ~\equiv~ \frac{ \| \underline{u}_{\mathrm{FLAME}} - \underline{u}_{\mathrm{mmc}} \|}
    { \| \underline{u}_{\mathrm{mmc}} \| }
\end{equation}
where $\underline{u}_{\mathrm{FLAME}}$ and $\underline{u}_{\mathrm{mmc}}$ are Euclidean
vectors of the nodal values of FLAME and mmc potentials,
respectively; the norms are Euclidean.

The dashed line in \figref{fig:error-vs-h-reg-grids-two-spheres} is a visual aid
indicating the $\mathcal{O}(h^2)$ slope and demonstrating quadratic convergence,
consistent with the number of basis functions (nine) used.

\subsubsection{Irregular stencils.}
\begin{figure}
\centering
  \includegraphics[width=0.75\linewidth]{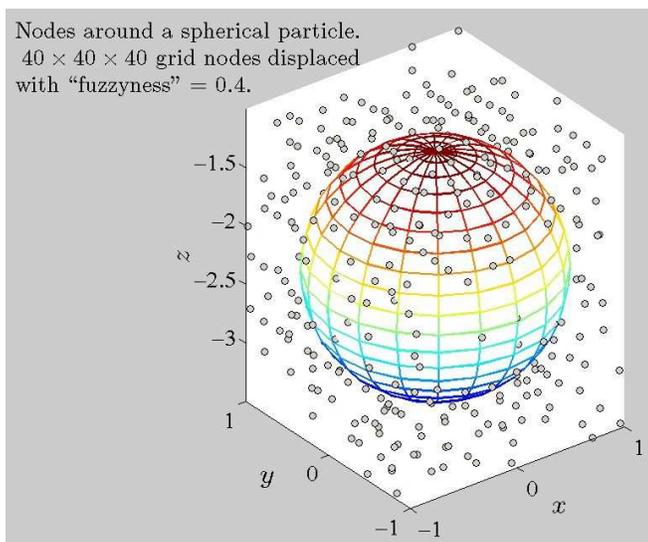}
  \caption{An irregular set of nodes for FLAME in the vicinity of a particle.
  The nodes are randomly displaced with respect to the regular Cartesian grid $40 \times 40 \times 40$,
  with the ``fuzziness'' (see text) of $f = 0.4$. Nodes inside the particle not visible
  but are distributed in a similar manner and with the same density.}
  \label{fig:fuzzy-grid-near-particle}
\end{figure}

%
A fragment of an irregular distribution of nodes near one of the two particles
is shown in \figref{fig:fuzzy-grid-near-particle}.
The random  displacements of the nodes are generated in the same manner as in the previous examples, except that
now these displacements are also applied in the $z$ direction in addition to $x$ and $y$.
The exact potential was imposed as the Dirichlet boundary condition for testing, to avoid any extraneous errors due to domain truncation.

\tableref{table:error-vs-fuzziness-3D} demonstrates the effect of ``fuzziness''
on the numerical accuracy. For all combinations of basis functions
and stencils in the table, the numerical accuracy deteriorates somewhat as the
``fuzziness'' $f$ increases. This is different from the previous 2D
example where the fuzziness was not a significant factor.

\begin{table}
  \centering
  \begin{tabular}{|c|c|c|c|c|c|}
    \hline
    Number of \\basis functions & Stencil size  & $f = 0$ & $f = 0.1$ & $f = 0.2$ & $f = 0.4$ \\
    \hline
    6	& 9	&  7.9E-04  & 8.05E-05 & 1.22E-04 & 2.69E-04 \\
    \hline
    6	& 19 &  4.72E-06 & 5.31E-06 & 1.06E-05 & 2.49E-05 \\
    \hline
    6	& 27 &   9.69E-06  & 9.17E-06	& 1.67E-05 & 1.74E-05 \\
    \hline
    9	& 19 &   2.13E-06  & 2.00E-06	& 3.10E-06 & 9.17E-06 \\
    \hline
    9	& 27 &  1.14E-05  & 1.12E-05	& 9.84E-06 & 6.68E-06 \\
    \hline
    \end{tabular}
  \caption{Relative errors in the nodal potential for FLAME on irregular grids. The reference quasi-analytical solution obtained by the multipole-multicenter (mmc) expansion with 64 terms per particle. The 3D test problem with two dielectric particles with the dielectric permittivity 5. $41^3$ nodes in FLAME.}\label{table:error-vs-fuzziness-3D}
\end{table}

The influence of the \emph{stencil size} for a fixed number of basis functions is less straightforward.
For six basis functions, the accuracy tends to improve when the stencil size changes
from 9 to 19 but deteriorates somewhat as the stencil expands further.

A plausible qualitative explanation is as follows. As the number of stencil nodes increases,
so does the robustness of the analytical approximation in FLAME,
because more information (in the form of the nodal values) is being utilized.
On the other hand, as the geometric size diam($\Omegarm^{(i)}$) of the stencil
``patch'' increases, the approximation accuracy, proportional to diam$(\Omegarm^{(i)})^p$,
deteriorates ($p$ is the order of approximation). These two conflicting trends lead to
the nontrivial dependence of the accuracy on the stencil size.
It would be worthwhile to analyze this rigorously in the future.

\subsubsection{Random nodes.}
\begin{figure}
\centering
  \includegraphics[width=0.8\linewidth]{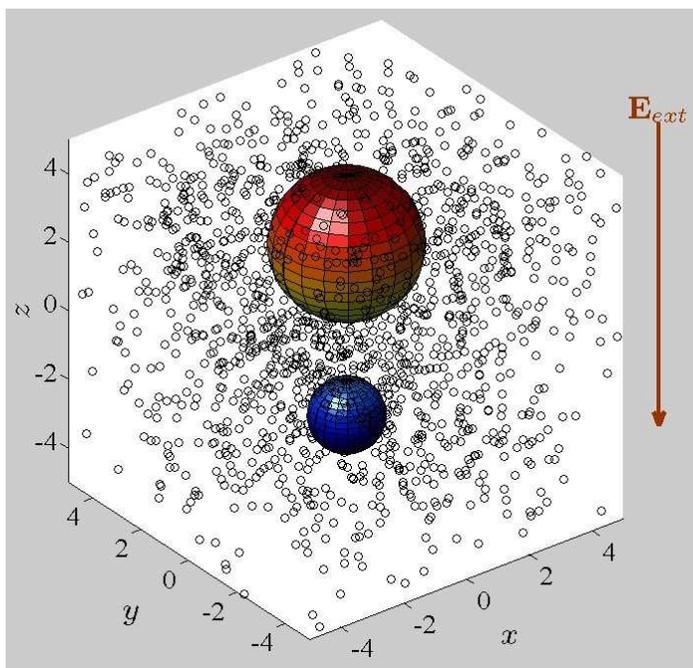}
  \caption{Two dielectric spheres in an external field. An example of
  least squares FLAME with 8,000 randomly generated nodes (only 1,600 shown for visual clarity).
  Nodes inside the particle not visible but are distributed with the same density.}
  \label{fig:two-spheres-3D}
\end{figure}

%
Further pushing the envelope, I have tested a completely random distribution of nodes
(as opposed to random displacements around a regular lattice), see \figref{fig:two-spheres-3D}. Clearly,
this can be viewed as the worst case scenario, and much better results
can be expected if sensible adaptive procedures \cite{Dai-adaptive-FLAME08}
are adopted in the future.



The initial test is performed with one particle only. In this case, since the exact solution
contains only the dipole harmonic and is thus included in the FLAME basis, one can expect
that FLAME will produce the exact solution up to the round-off error. This has been confirmed
experimentally. For example, for 5000 FLAME nodes and the 27-point FLAME scheme with 16 basis functions,
the relative error in the potential at the nodes was on the order of $10^{-13}$, despite the fact
that the distribution of the nodes is random and hence carries no geometric information whatsoever.

Note that 5000 nodes correspond to only $\sim 17$ points per coordinate, with the average separation distance
of $h \approx L_{\mathrm{dom}} / 17 \approx 0.59$, which is comparable with the radius of the particle.
Clearly, classical FD schemes would require much finer meshes for high accuracy
(and still would not be able to deliver machine precision).

\begin{figure}
\centering
  \includegraphics[width=0.8\linewidth]{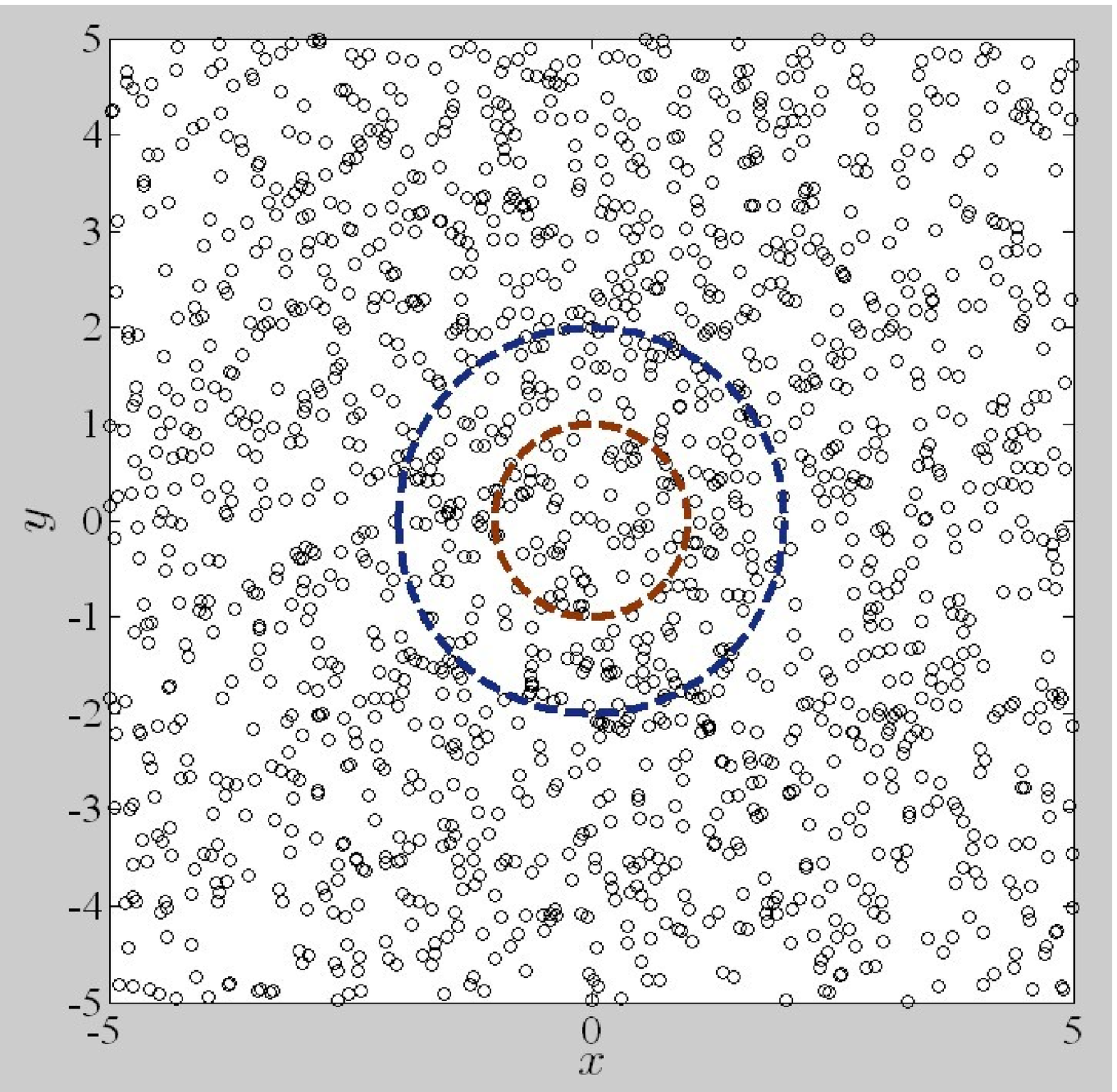}
  \caption{Same as in the previous figure, but projections of the nodes
  and of the two particles on the $xy$-plane shown. Note that the exterior Dirichlet
  boundary in this example is ``fuzzy''.}
  \label{fig:two-spheres-xy-proj}
\end{figure}

Let us now return to the test problem with \emph{two particles}, as described above.
\figref{fig:two-spheres-3D} presents the general setup of the problem solved with a random distribution of 8000 nodes (only 1600 shown in the picture); the simulation results are reported below.
\figref{fig:two-spheres-xy-proj} shows projections of the same nodes and of the spheres on the $xy$-plane.

\tableref{table:error-random-nodes-vs-n-3D} shows the dependence of
the numerical error on the number of random nodes. The errors are still quite
small despite the highly suboptimal distribution of nodes. At the same time,
as could be expected, the errors are higher than for regular grids and,
moreover, convergence of the solution with respect to the number of nodes
is not monotone. This is almost certainly attributable to ``poor quality''
stencils, although precise \emph{a priori} measures of quality in FLAME are still to be developed;
see further comments on that below.

\begin{table}
  \centering
  \begin{tabular}{|c|c|c|}
    \hline
    Number of nodes & Effective ``grid size'' & Relative error \\
    \hline
    10000	& 0.464	&	1.94E-04\\
    \hline
    20000	& 0.368	&   3.66E-04\\
    \hline
    30000	& 0.322	&	2.76E-05\\
    \hline
    40000	& 0.292	&	3.16E-05 \\
    \hline
    60000 &  0.255  &  2.11E-05 \\
    \hline
  \end{tabular}
  \caption{Relative errors in the nodal potential for FLAME on \emph{fully random} sets of nodes.
  The reference quasi-analytical solution is obtained by the multipole-multicenter (mmc)
  expansion with 64 terms per particle.
  3D test problem with two dielectric particles with the dielectric permittivity 5.
  19-point stencils with 9 basis functions.}\label{table:error-random-nodes-vs-n-3D}
\end{table}

One may envision that in future applications of the new version of FLAME the nodal distributions will be optimized and generated adaptively, based on appropriate \emph{a posteriori} error estimates.
Such distributions can then be expected to produce more accurate solutions than
random, semi-random or even regular sets of modes would.

A natural \emph{a posteriori} error indicator\footnote{As in FEM, one may want to
distinguish \emph{estimators} that produce some quantitative measure of the error
from qualitative \emph{indicators} of regions where grid refinement is desirable.}
in FLAME is the discrepancy between the values of the solution over two overlapping
patches; this indicator has already been shown to produce quite reasonable results,
leading to grid refinement around the gaps between two neighboring particles \cite{Dai-adaptive-FLAME08,Dai-Webb09}.
It should certainly be possible to develop other, and better, indicators in the future.

Likewise -- and also in parallel with FEM where similar issues have been studied for several decades --
good \emph{a priori} error estimates indicating the ``quality'' of irregular stencils are needed.
In FEM, there are several geometric conditions that characterize the quality of elements
(``quality'' being ultimately understood as approximation accuracy):
for example, the minimum or maximum angle conditions, the radius of the inscribed sphere
vs. the element diameter, etc.

In comparison with FEM, irregular stencils (especially large ones) are difficult to characterize
in pure geometric terms. Therefore particularly relevant are the \emph{algebraic} conditions.
In FEM, such conditions are related to the properties of the affine transformation
of a given element to the canonical ``master'' element \cite{AlShenk94,Ciarlet78}
or, in less traditional analysis (see \cite{Tsukerman-book07}, \textsection 3.14), to the  maximum eigenvalue
of the element stiffness matrix or the minimum singular value of the edge shape matrix
\cite{Tsukerman-general-criterion98,Tsukerman-appeox-conservative-fields98,Tsukerman-Plaks-comparison-criteria98,Tsukerman-Plaks-approx-errors99}.
One conjecture that may form an interesting subject for future research is that for least squares FLAME the minimum singular value of the nodal matrices such as $N$ \eqnref{eqn:N-matrix} will play a similar role.
%
\section{Conclusion}
\label{sec:Conclusion}
%
FLAME replaces standard Taylor expansions of classical finite difference methods
with local approximations of the solution by functions satisfying the
underlying differential equation and interface boundary conditions. As a result,
it is not unusual for the consistency error of FLAME to be orders of magnitude
lower than that of conventional schemes, due to the high accuracy of the Trefftz
approximations employed in FLAME.

Previously one constraint in FLAME has been that the number of approximating functions
had to be closely linked to the number of nodes in the grid stencil. The present paper,
by expanding on the idea of Pinheiro \& Webb \cite{Pinheiro09}, removes
this constraint: while Trefftz functions still provide very accurate approximation,
the numbers of functions and nodes are now decoupled. This is accomplished
by using least-squares matching of the approximate solution at the stencil nodes.
Consequently, FLAME schemes can now be generated on irregular stencils with the number of nodes
substantially greater than the number of approximating functions. This extends
the practical applicability of the method.

Counterintuitively, the use of the least squares fit instead of the exact nodal match
does not reduce the numerical accuracy; on the contrary, the method tends to become more robust.
The consistency error is still governed primarily by the approximation accuracy
of the solution with the Trefftz-FLAME basis.

In contrast with meshless methods and other similar techniques,
the treatment of interface and exterior boundary conditions in FLAME is simple and natural.
Trefftz-FLAME basis functions satisfy the underlying equation and
the boundary conditions by construction and therefore do not require any additional
constraints to be imposed.

In summary, the new version of FLAME is a promising technique that
combines the high accuracy of Trefftz approximations with robust operation
on irregular stencils.
Once a set of basis functions is established, FLAME is exceptionally easy
to implement in computer codes, as it does not require mesh generation,
numerical integration or differentiation. The ability of the method to handle irregular stencils,
as well as the \emph{a posteriori} error indicators inherent in FLAME,
bodes well for the development of adaptive algorithms in the future.

The price to pay for these advantages is the need to derive the local Trefftz sets.
Expanding the library of canonical solutions (e.g. elliptic problems / wave scattering;
(piecewise) planar / cylindrical / spherical boundaries, etc.)
may lead to the commensurate expansion of the applicability of FLAME.
On the other hand, problems where local solutions are unavailable or difficult
to obtain may remain out of reach for FLAME until additional ideas
are put forward.

An interesting subject for future research
is \emph{a priori} and \emph{a posteriori} error estimates in FLAME.
Some considerations are already presented in this paper.

In comparison with the previous version of Trefftz-FLAME,
least squares FLAME is particularly promising in cases where
an intrinsic disparity between the natural sizes of the bases and stencils exists.
Examples include finite difference time domain simulation of wave propagation,
as well as electromagnetic vector problems where this version of FLAME
has already been explored \cite{Pinheiro09}.

\end{document}